\documentclass[preprint,aps,showpacs]{revtex4}       
\usepackage{graphicx}
\begin{document}

\title{Coupled-channels calculations of $^{16}$O+$^{16}$O fusion}
\author{H. Esbensen}
\affiliation{Physics Division, Argonne National Laboratory, Argonne, Illinois 60439}
\date{\today}

\begin{abstract}
Fusion data for $^{16}$O+$^{16}$O are analyzed by coupled-channels 
calculations. It is shown that the calculated cross sections are 
sensitive to the couplings to the $2^+$ and $3^-$ excitation channels 
even at low energies, where these channels are closed. 
The sensitivity to the ion-ion potential is investigated by applying 
a conventional Woods-Saxon potential and the M3Y+repulsion potential, 
consisting of the M3Y double-folding potential and a repulsive 
term that simulates the effect of the nuclear incompressibility.
The best overall fit to the data is obtained with a M3Y+repulsion
potential which produces a shallow potential in the entrance channel.
The stepwise increase in measured fusion cross sections at high 
energies is also consistent with such a shallow potential. 
The steps are correlated with overcoming the barriers for the 
angular momenta $L$ = 12, 14, 16, and 18.
To improve the fit to the low-energy data requires a shallower 
potential and this causes a even stronger hindrance of fusion at low 
energies. It is therefore difficult, based on the existing fusion data, 
to make an accurate extrapolation to energies that are of interest to
astrophysics.
\end{abstract}
\maketitle

\section{Introduction}

A major challenge in nuclear astrophysics is to measure the cross sections 
for radiative capture and fusion reactions with high precision and down
to very low energies. Another challenge is to develop models that can
reproduce the existing data and be used with confidence to extrapolate 
the cross sections to the energies that are of interest to astrophysics. 
Examples of reactions where these challenges exist are the low-energy 
fusion of $^{12}$C+$^{12}$C, $^{12}$C+$^{16}$O, and $^{16}$O+$^{16}$O.
It was recently suggested \cite{cljastro} that the fusion rates that have 
been used in the past for these reactions should be reduced because of
a hindrance phenomenon, and the implications for stellar
burning and nucleosynthesis were investigated in Ref. \cite{gasques}.
The fusion hindrance has been observed experimentally at
extreme subbarrier energies in  many medium-heavy systems \cite{cljsys} 
but it has not yet been observed convincingly in lighter systems.
It is therefore of interest to study the possible evidence for such a
phenomenon in light systems.

The theoretical description of light-ion fusion reactions,
for example, of the $^{16}$O+$^{16}$O fusion data \cite{Thomas}, 
is often limited to optical model calculations.
An exception is the study by Reinhard et al. \cite{PGR} in which the 
ion-ion potential for $^{16}$O+$^{16}$O was derived from an adiabatic 
TDHF calculation, and the fusion cross sections that were calculated 
have served as guidance for the extrapolation to low energies. 
The purpose of this work is to use the coupled-channels method to analyze 
the $^{16}$O+$^{16}$O fusion data of Ref. \cite{Thomas} in order to 
investigate the influence of the couplings to the $2^+$ and $3^-$ 
excitations of the reacting nuclei
and to see whether the fusion hindrance phenomenon is 
likely to exist in such a light system.

The coupled-channels method has been used in numerous analyses
of the fusion data for medium-heavy nuclei. It has been very 
successful in many cases in reproducing the data at energies 
near and slightly below the Coulomb barrier, typically 
down to 0.1 mb or 0.01 mb.
Thus it has been possible to generate the large enhancement that is needed
to fit the data at subbarrier energies by including couplings to surface 
excitation modes, primarily to the low-lying $2^+$ and $3^-$ states, and 
to two-phonon and mutual excitations of these states
(see, for example, the review article Ref. \cite{baha} and the proceedings 
Refs. \cite{fus03,fus06}.) 
In some cases it is necessary to include couplings to transfer channels
(see Refs. \cite{fus03,fus06}.)

The definition of fusion in the coupled-channels approach is usually based
on ingoing wave boundary conditions (IWBC). They are imposed at a distance 
somewhere inside the Coulomb barrier, for example, at the minimum of the pocket 
in the entrance channel potential. However, a short-ranged imaginary potential 
has also been used to simulate the fusion. The two ways of defining the fusion 
give essentially the same result in most cases. This supports the view that 
the fusion is primarily sensitive to the description in the vicinity of and 
outside the Coulomb barrier, and the empirical proximity type
Woods-Saxon potential \cite{AW}, which is based on extensive analyses of
elastic scattering data and also on the M3Y double-folding potential, has 
served as a very realistic interaction in coupled-channels calculations.

The above view of the fusion process has been challenged by the discovery 
of the fusion hindrance at extreme subbarrier energies.
Since the hindrance sets in at a rather high excitation energy of the 
compound nucleus, is was suggested early on that it had to be an entrance
channel phenomenon \cite{niy}. It was shown in Ref. \cite{dasso} that the 
hindrance could be explained by adjusting the ion-ion potential at small
distances between the reacting nuclei so that it produced a shallow potential 
in the entrance channel. In contrast, the empirical interaction of 
Ref. \cite{AW} produces a relative deep pocket, and the M3Y double-folding
potential is unphysical because it produces a pocket that is much deeper
than the ground state energy of the compound nucleus. It was proposed in Ref.
\cite{dasso} that the new low-energy fusion data offer the opportunity to 
investigate the radial dependence of the ion-ion potential at short distances.

We have shown that a shallow potential can be constructed by correcting 
the M3Y double-folding potential with a repulsive term that simulates the 
effect of the nuclear compressibility \cite{misi2}.  
Thus we were able to explain successfully 
the fusion data for $^{64}$Ni+$^{64}$Ni \cite{misi1}, $^{28}$Si+$^{64}$Ni 
\cite{sini}, and $^{16}$O+$^{208}$Pb \cite{opb}, ranging in cross sections 
from 20 nb and up to 1 b. It is therefore of interest to apply this type of 
potential in the analysis of the $^{16}$O+$^{16}$O fusion data, in order
to see how it will affect the extrapolation to extreme subbarrier energies.

With an increased sensitivity (at extreme subbarrier energies) to the potential
at small distances between the fusing nuclei, one should also be concerned
about the validity of the basic assumptions in the coupled-channels approach, 
namely, that the structure input is that of the isolated nuclei. 
For strongly overlapping nuclei, the modes of excitations may be better 
described in terms of excitations of the compound nucleus. Unfortunately, 
it is very difficult in the coupled-channels approach to develop a model
that includes a realistic transition from a di-nuclear description 
to a compound nucleus description. 
Some justification for the coupled-channels method can be found, for example,
in calculations that are based on the two-center Hartree-Fock method. 
Thus, in the case of $^{16}$O+$^{16}$O, Zint and Mosel found that 
the individual shell structures of the two $^{16}$O nuclei survive up 
to a remarkable degree of overlap \cite{mosel}. Moreover, the ion-ion
potential they determined gives a very shallow potential in the entrance 
channel, in qualitative agreement with the empirical findings from the
analysis of elastic scattering data \cite{gobbi}, and also, as we shall
see, from the analysis of the fusion data.

\section{Coupled-channels description}

The fusion cross sections for the oxygen isotopes are somewhat exceptional 
in the sense that they are rather structureless \cite{Thomas}, whereas 
fusion data for $^{12}$C+$^{12}$C \cite{Aguilera} and $^{12}$C+$^{16}$O 
\cite{Patterson} contain very rich structures or resonances.
Since the coupled-channels calculations presented here produce rather 
smooth and structureless cross sections at low energies, the 
investigations will be restricted to the fusion of $^{16}$O+$^{16}$O.

The basic assumptions and ingredients in the coupled-channels description 
of fusion reactions are summarized below.
The structure input for the $2^+$ and $3^-$ states in $^{16}$O 
\cite{ENSDF,alfa} is shown in Table I.  The channels for the 
two excited states in $^{16}$O become closed when the center-of-mass 
energy is less than the excitation energy, $E_x$ $\approx$ 6-7 MeV. 
The asymptotic boundary condition for the radial wave function $u_{nL}(r)$
in an open, inelastic channel $n$ with angular momentum $L$ is the 
out-going Coulomb wave,
\begin{equation}
u_{nL}(r) \propto O_L(q_nr) = G_L(\eta_n,q_nr) + i F_L(\eta_n,q_nr), \ \ \ {\rm for} \ \ 
r\rightarrow\infty,
\label{ow}
\end{equation}
where $\hbar q_n$
is the asymptotic relative momentum and $\eta_n=Z_1Z_2e^2/(\hbar v)$ is the 
Sommerfeld parameter for channel $n$.
The out-going wave is here expressed in terms of the regular and 
irregular Coulomb wave functions $F_L$ and $G_L$, respectively.  
For a closed channel the condition (\ref{ow}) is replaced by
\begin{equation}
u_{nL}(r) \propto W_{-\eta_n,L+1/2}(2q_nr),
\end{equation}
where $W_{-\eta,L+1/2}(z)$ is the Whittaker function.

The coupled equations are solved in the so-called rotating frame 
approximation. 
A detailed discussion of the coupled-equations
in this approximation can be found, for example, in Ref. \cite{hes03}.
The boundary conditions that are used at short distances 
are the ingoing-wave boundary conditions (IWBC), which are imposed at the 
location of the minimum of the pocket in the entrance channel, and the 
fusion cross section is determined by the ingoing flux. This model works 
quite well at energies that are near and below the Coulomb barrier.

It is a common problem that one cannot always reproduce the fusion data at 
extreme subbarrier energies and at energies far above the Coulomb barrier 
by using exactly the same model assumptions in the two energy regimes. 
It has been suggested that the problem can be solved by considering 
the effect of decoherence \cite{diaz} but calculations that
demonstrate this point were not carried out.
The solution we have used \cite{opb} is to  supplement the IWBC at 
energies far above the Coulomb barrier with a weak imaginary potential 
that acts near the minimum of the pocket in the entrance channel.
The need for such an imaginary potential at high energies may reflect 
the influence of an increasing number of reaction channels, which cannot 
be considered explicitly in a practical calculation. 

\subsection{Standard Woods-Saxon potential}

The real part of the ion-ion potential is commonly parametrized as
a Woods-Saxon potential, 
\begin{equation}
V(r) = \frac{V_0}{1+\exp[(r-R_{\rm pot})/a]},
\end{equation}
and the proximity type potential discussed in Ref. \cite{AW} 
(Eq. (40) of Section III.1) will be used below.  
The results of coupled-channels calculations that are based
on this potential will be compared to data and to calculations that use
the M3Y+repulsion potential \cite{misi2}.
These two types of potentials are basically identical at large radial distances
between the reacting nuclei, and they produce essentially the same Coulomb 
barrier height. They differ at short distances, as will be shown in the next 
section. The Woods-Saxon potential produces a relatively deep pocket 
in the entrance channel potential, whereas the M3Y+repulsion potential can be 
adjusted to produce a shallow pocket and a thicker Coulomb barrier. 
The latter two features help explain the hindrance of fusion 
\cite{misi1,misi2,sini,opb}, which has been observed in many heavy-ion systems 
at extreme subbarrier energies.  The issue here is whether an analysis of the 
fusion data for $^{16}$O+$^{16}$O will show a sensitivity to the potential
at short distances.

\subsection{Fusion cross sections}

\begin{figure}
\includegraphics[width = 12cm]{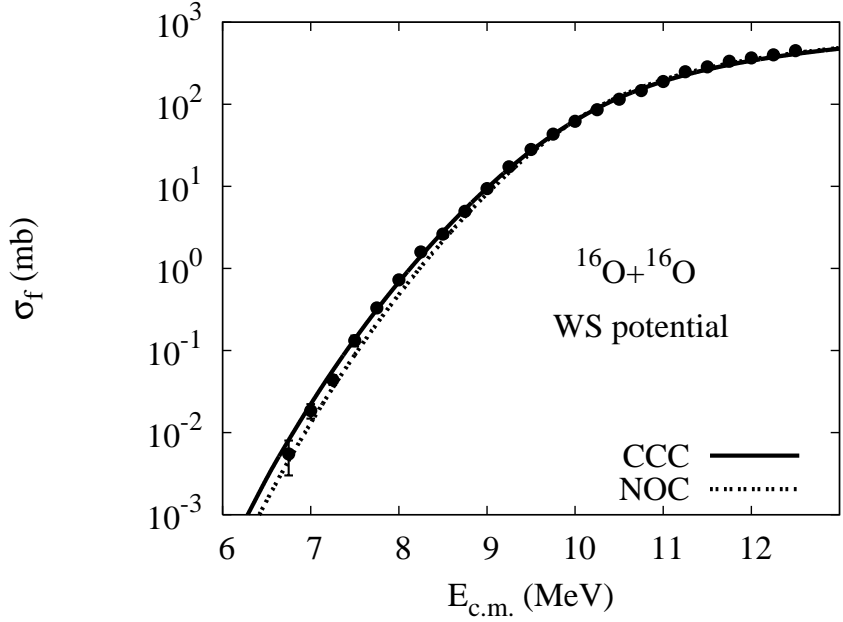}
\caption{\label{fus} The fusion cross sections for $^{16}$O+$^{16}$O 
measured by Thomas et al. \cite{Thomas} are compared to coupled-channels 
calculations (CCC) and to the no-coupling limit (NOC). 
Both calculations are based on the Woods-Saxon potential
described in the text.}
\end{figure}

The fusion cross sections for $^{16}$O+$^{16}$O that were measured 
by Thomas et al. \cite{Thomas} are compared in Fig. \ref{fus} to
coupled-channels calculations and to the no-coupling limit,
i.~e., a one-dimensional barrier penetration calculation. 
Both calculations are based on the standard 
proximity type, Woods-Saxon potential \cite{AW}.
The uncertainty in the data was taken from the published figures, 
except at center-of-mass energies larger than 8 MeV, where an 
(arbitrary) uncertainty of 5\% was adopted, because it was 
not possible to read the small experimental uncertainty. 
Earlier measurements do exist, see for example Refs. \cite{Spinka,Hulke,Wu}, 
but they will not be shown here.

The proximity type, Woods-Saxon potential \cite{AW}, in which the 
radius has been adjusted to provide the best fit to 
the $^{16}$O+$^{16}$O fusion data \cite{Thomas}, has the parameters: 
$V_0$ = -42.14 MeV, $R_{\rm pot}$= 6.083 fm, and $a$ = 0.602 fm.
The solid curve (CCC) in Fig. \ref{fus} is the coupled-channels result 
one obtains with this potential and it has a  $\chi^2/N$ = 1.5.
The dashed curve shows the no-coupling limit (NOC) one obtains with 
the same potential.  The fit to the data in the no-coupling limit
can be improved by adjusting the radius of the potential. 
The best fit is achieved for $R_{\rm pot}$ = 6.133 fm and has a 
$\chi^2/N$ = 4.2.
This is much larger than the $\chi^2/N$ = 1.5 obtained in the 
coupled-channels calculation and shows that the couplings to the 
$2^+$ and $3^-$ states do play a significant role.

It may be seen in Fig. \ref{fus} that the measured cross sections fall off 
faster with decreasing energy than predicted by the coupled-channels
calculation at the lowest energies. This is a signature of the onset of 
the fusion hindrance phenomenon discussed earlier. The hindrance will 
be explored further in the following sections.

\section{The M3Y+repulsion potential}

The calculation of the M3Y+repulsion potential is described in 
Ref. \cite{misi2} but some of the essential features are summarized
here. First one calculates the M3Y double-folding potential
(including the exchange term). This requires as input the 
proton and neutron densities of projectile and target. 
The densities of protons and neutrons were assumed to be identical 
for $^{16}$O and the form of density was assumed to be a fermi function 
with radius $R$=2.5 fm and diffuseness $a$=0.52 fm. 
The radius was adjusted so that the measured RMS charge radius 
of 2.737(8) fm \cite{devries} was reproduced.

The repulsive term associated with the nuclear incompressibility
is also obtained from the double-folding procedure using a 
repulsive effective NN interaction of the form $V_{\rm rep}\delta({\bf r})$.
The density that is used in connection with the repulsive term 
has the same radius as the ordinary densities mentioned above 
but the diffuseness $a_{\rm rep}$ is chosen differently.
Thus there are two parameters in the calculation of the repulsive 
interaction, the strength $V_{\rm rep}$ and the diffuseness $a_{\rm rep}$
of the density. They are constrained, as explained in Ref. \cite{misi2},
so that the total nuclear interaction for completely overlapping nuclei,
$U_N(r=0)$, is consistent the equation of state $\epsilon(\rho)$
at normal nuclear matter density $\rho$, and $\epsilon(2\rho)$
at twice the nuclear matter density.
This condition was expressed in Ref. \cite{misi2} by the relation
\begin{equation}
U_N(r=0) = 2A_a\Bigl(\epsilon(2\rho)-\epsilon(\rho)\Bigr)
\approx \frac{A_a}{9} \ K.
\end{equation}
Here $A_a$ is the mass number of the smaller nucleus, so the equation
expresses the change in energy one has by embedding the smaller nucleus 
inside the larger. The last approximation relates this change in energy 
to the nuclear incompressibility, $K=9\rho^2 (d^2\epsilon(\rho)/d\rho^2)$.

\begin{figure}
\includegraphics[width = 12cm]{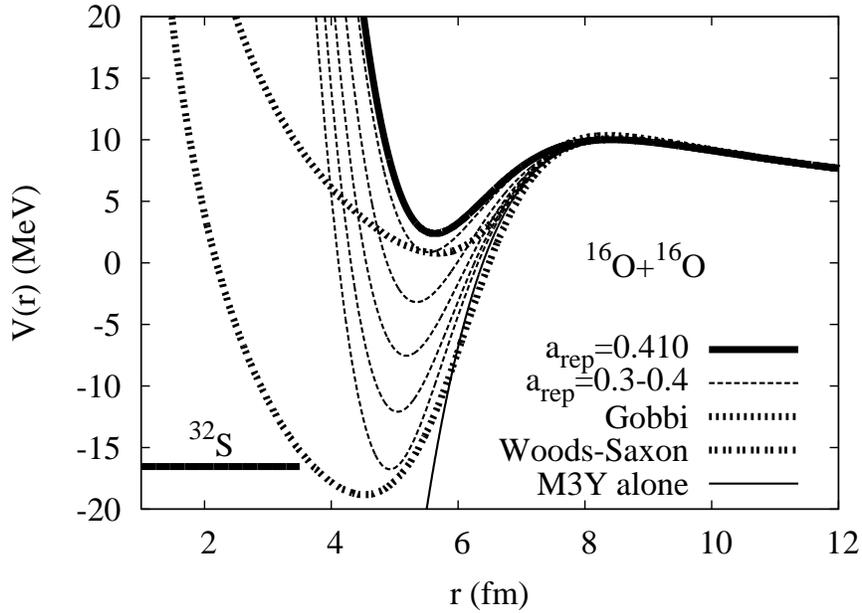}
\caption{\label{pot} The M3Y+repulsion entrance-channel potential is
shown for $a_{\rm rep}$=0.41 fm, and for $a_{\rm rep}$ = 0.3-0.4 fm 
in steps of 0.025 fm.
The Woods-Saxon, the Gobbi \cite{gobbi}, and the pure M3Y entrance channel 
potentials are also shown, and the energy of the compound nucleus $^{32}$S
is indicated.}
\end{figure}

The entrance channel potential obtained from the M3Y+repulsion potential 
is illustrated in Fig. \ref{pot} for a range of values of 
the diffuseness parameter $a_{\rm rep}$. The strength of the repulsive
interaction $V_{\rm rep}$ was adjusted in each case to produce the
nuclear incompressibility $K$ = 234 MeV. This is the value
that has been obtained from the Thomas-Fermi equation of state 
for symmetric nuclear matter \cite{myers}.
The smallest value of $a_{\rm rep}$, which is 0.3 fm, produces a pocket 
that is as deep as the energy of the compound nucleus $^{32}$S.  
The largest value, $a_{\rm rep}$=0.41 fm, produces a pocket at 2.4 MeV
and has a Coulomb barrier of 10.01 MeV.

The thin solid curve in Fig. \ref{pot} is the entrance channel potential
one obtains with the pure M3Y potential (including the exchange term).
It has an unrealistic and extremely deep pocket, which is far below
the energy of the compound nucleus. 
The entrance potential, which is based on the Woods-Saxon potential
discussed in the previous section, is shown by the lower thick dashed 
curve. It is slightly deeper than the ground state of the compound 
nucleus and it has a Coulomb barrier of 10.10 MeV. 

Finally, the upper thick dashed curve in Fig. \ref{pot} is the shallow 
Gobbi potential \cite{gobbi}, which, by the way, is in surprisingly good 
agreement with the potential obtained in the two-center Hartree-Fock 
calculation of Ref. \cite{mosel}.
It is seen that the M3Y+repulsion and Gobbi potentials have almost the
same depth but the thickness of the Coulomb barrier is different. 
It turns out that the thicker barrier provided by the M3Y+repulsion 
potential gives a much better fit to the fusion data when applied in 
the coupled-channels calculations.

\subsection{Fusion cross sections}

\begin{figure}
\includegraphics[width = 12cm]{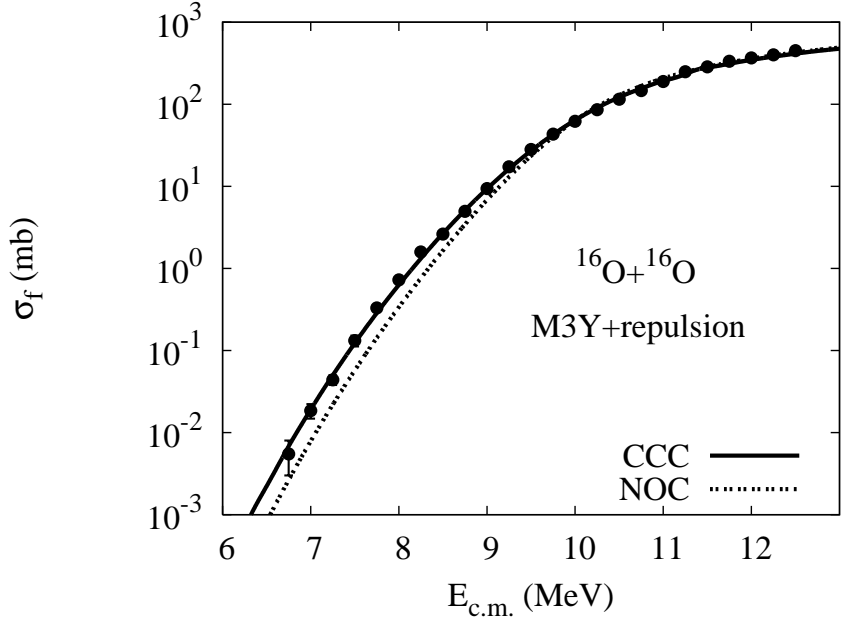}
\caption{\label{fm3y} The fusion data for $^{16}$O+$^{16}$O \cite{Thomas} 
are compared to coupled-channels calculations (CCC) and to the no-coupling 
limit (NOC). Both calculations are based on the M3Y+repulsion potential 
with $a_{\rm rep}$ = 0.41 fm.}
\end{figure}

The thick solid curve in Fig. \ref{pot}, which is based on the diffuseness
parameter $a_{\rm rep}$ = 0.41 fm, is the entrance channel potential that
provides the best fit to the fusion data in the coupled-channels calculations.
The calculated cross section is shown by the solid curve in Fig. \ref{fm3y}
and it has a $\chi^2/N$ = 1.3.  The quality of the fit is only slightly 
better than what was obtained in the previous section using the standard 
Woods-Saxon potential. The slight improvement is difficult to see but it 
is achieved mainly at the lowest energies.

\begin{figure}
\includegraphics[width = 12cm]{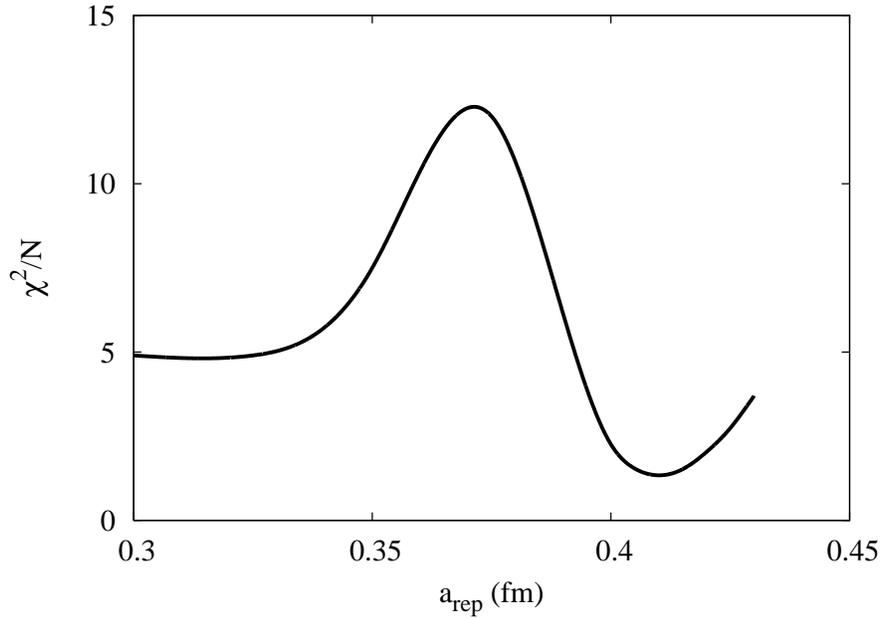}
\caption{\label{chi} The $\chi^2/N$ obtained from the $^{16}$O+$^{16}$O 
fusion data \cite{Thomas} and coupled-channels calculations.
The $\chi^2/N$  is shown as function of the diffuseness parameter 
$a_{\rm rep}$, which determines the repulsive term in the 
M3Y+repulsion potential.}
\end{figure}

The $\chi^2/N$ is shown in Fig. \ref{chi} as function of the 
diffuseness parameter $a_{\rm rep}$. There are two minima, one at a small 
value, $a_{\rm rep}\approx$ 0.325, and one at $a_{\rm rep}$ = 0.41 fm, 
which is by far the best solution.
It is of interest to compare $a_{\rm rep}$ to the values that have been 
used for other systems.  Thus for the $^{64}$Ni+$^{64}$Ni system we 
obtained the best fit to the data  for $a_{\rm rep}$ = 0.403 fm \cite{misi2}. 
For $^{28}$Si+$^{64}$Ni the value was 0.392 fm \cite{sini},
and for the very asymmetric system $^{16}$O+$^{208}$Pb 
we had to use the smaller value $a_{\rm rep}$ = 0.35 fm \cite{opb}.

\begin{figure}
\includegraphics[width = 12cm]{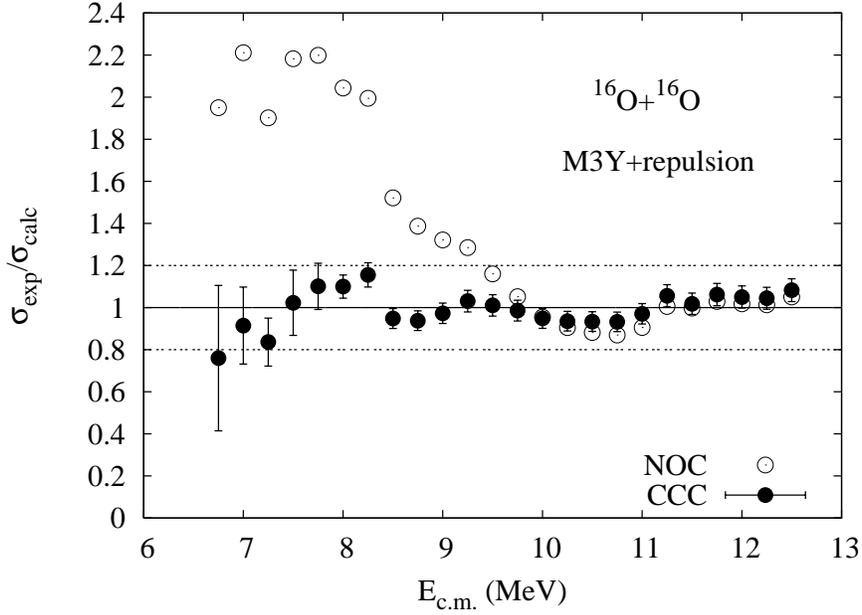}
\caption{\label{rm3y} Ratio of experimental \cite{Thomas} and 
calculated cross sections shown in Fig. \ref{fm3y}.}
\end{figure}

The discrepancy with the data is emphasized in Fig. \ref{rm3y} where 
ratios of the measured and calculated fusion cross sections are shown.
The solid circles are the coupled-channels results and the open
circles show the results in the no-coupling limit obtained with the 
same potential, namely, the M3Y+repulsion potential with 
$a_{\rm rep}$ = 0.41 fm.
The effect of the couplings to the $2^+$ and $3^-$ excitations is 
to bring the cross section ratio closer to one.
However, there are still some minor deviations from one. For example,
the ratio of the measurement and the coupled-channels calculation 
(solid circles) shows a decreasing trend with decreasing energy
below 8 MeV and it is less than one at the lowest energy point. 
This is a signature of the experimental fusion hindrance with 
respect to the coupled-channels calculation.

\section{The $S$ factor at low energies}

It is unfortunate that the quality of the fit of the coupled-channels
calculations to the $^{16}$O+$^{16}$O fusion data is essentially the 
same whether we use the Woods-Saxon or the M3Y+repulsion potentials. 
From the empirical knowledge of the fusion hindrance phenomenon 
\cite{cljsys} one would have expected that the M3Y+repulsion potential 
would provide a much better description of the low-energy data.
However, the improvement is modest. One would need measurements 
at even lower energies in order to be able to see a stronger 
sensitivity to the ion-ion potential at short distances.

\begin{figure}
\includegraphics[width = 11cm]{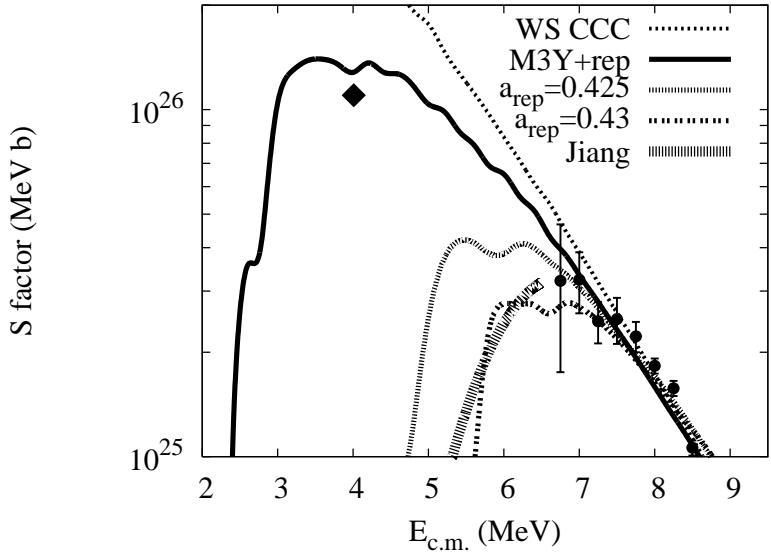}
\caption{\label{sf} $S$ factors for the coupled-channels calculations
shown in Figs. \ref{fus} and \ref{fm3y} are compared to the data 
\cite{Thomas}. Also shown are coupled-channels calculations
that are based on shallower potentials, with $a_{\rm rep}$ = 0.425 and
0.43 fm, respectively.
The diamond is the adiabatic TDHF results \cite{PGR}.  The lowest thick 
curve is the extrapolation made by Jiang et al. \cite{cljastro}.}
\end{figure}

A good way to emphasize the low-energy behavior of the measured and 
calculated fusion cross sections is to plot the $S$ factor for fusion
defined by 
\begin{equation}
S = E_{c.m.} \sigma_f \ \exp(2\pi\eta),
\end{equation}
where $\eta$ 
is the Sommerfeld parameter.
The experimental $S$ factors are compared in Fig. \ref{sf} to the two 
coupled-channels calculations that were discussed earlier. The top dashed 
curve is based on the Woods-Saxon potential, whereas the solid curve is 
based on the M3Y+repulsion potential. The latter provides a slightly 
better fit to the data at the lowest energies but the error bars are 
large so the overall improvement in terms of a $\chi^2/N$ is modest, 
as discussed in the previous section.

The two coupled-channels calculations shown in Fig. \ref{sf} start to 
deviate as the energy is reduced.  The calculation that is based on the 
M3Y+repulsion potential (the solid curve) develops a maximum near 4 MeV.
The reason is that the entrance channel potential has a pocket at 2.4 MeV, 
and this forces the $S$ factor to vanish below that energy when the fusion 
is determined by IWBC. 
It is interesting that the value of the $S$ factor at 4 MeV (solid curve)
is in fair agreement with the prediction of the adiabatic TDHF calculation 
\cite{PGR}. However, the $S$ factor for fusion obtained in the adiabatic 
TDHF calculation keeps increasing with decreasing energy \cite{PGR}.  

It is not clear a priori whether the $S$ factor for the fusion of
$^{16}$O+$^{16}$O should develop a maximum at low energy. 
It does not have to do that because the ground state Q value for 
producing $^{32}$S is positive.  It is only for negative Q-values 
one can argue that the $S$ factor must have a maximum at some positive 
center-of-mass energy \cite{cljsys}. 

There is an alternative extrapolation method \cite{cljsys}
which is based on the logarithmic derivative 
of the energy-weighted fusion cross section, 
\begin{equation}
L(E)
= \frac{1}{E_{c.m.}\sigma_f} \frac{d(E_{c.m.}\sigma_f)}{dE}.
\end{equation}
This quantity has a nearly linear dependence on energy at extreme
subbarrier energies in most of the medium-heavy systems that have 
been studied experimentally. The linear dependence makes it fairly 
easy to extrapolate the data to the energy where the $S$ factor 
has a maximum \cite{cljsys}. 

A better parametrization of $L(E)$ was adopted in Ref. \cite{cljastro}.  
By considering all of the 50 data points that have been measured
below 8.5 MeV \cite{Wu,Spinka,Hulke,Thomas} it was 
concluded that the $S$ factor for the fusion of $^{16}$O+$^{16}$O must 
have a maximum close to 7 MeV. 
The extrapolation to lower energies that was obtained in Ref. \cite{cljastro} 
is shown in Fig. \ref{sf} in terms of the $S$ factor by the lowest, thick curve. 
The low energy fusion cross sections predicted by this extrapolation 
are even more suppressed than the coupled-channels calculation that
is based on the M3Y+repulsion potential.

Apparently, there are certain features of the data that are not 
reproduced by the coupled-channels calculations presented here. 
Some indications of that can be seen in the cross section ratios
shown in Fig.  \ref{rm3y}. For example, the seven lowest data points 
form an isolated group which is disconnected from the rest above 8.5 MeV.
It is not clear what causes the discontinuity;
is it a remnant of a resonance or is it an experimental problem? 
In any case, one can adjust the M3Y+repulsion potential so that 
the coupled-channels calculations reproduce the the energy 
dependence of the seven lowest points, i.~e., so that the cross 
section ratio $\sigma_{exp}/\sigma_{calc}$ becomes a constant.
This can be achieved  with a diffuseness parameter in the range 
$a_{\rm rep}$ = 0.425 - 0.43 fm, which produces a  pocket
in the entrance channel potential in the range of 4.5 to 5.2 MeV.
The $S$ factors obtained from such calculations are also shown 
in Fig. \ref{sf}. The results are in fair agreement with the 
extrapolation method proposed by Jiang et al. \cite{cljastro}.
That is not surprising because the latter extrapolation 
was also based on low-energy data.

Fusion should in principle be allowed down to zero energy because 
the ground state of the compound nucleus $^{32}$S is at a much lower
energy (see Fig. \ref{pot}). To describe the fusion at such low
energies would require an extension of the model used here, 
for example along the lines proposed in Ref. \cite{ichikawa}.

\section{Fusion at high energies}

Another way to test the ion-ion potential is to compare to cross
sections that have been measured at energies far above the Coulomb barrier.
Here the data are often suppressed compared to calculations
that are based on a conventional Woods-Saxon potential,
with a relatively deep entrance potential \cite{dasg}.
We have previously shown that the shallow entrance channel potential,
produced by the M3Y+repulsion potential, gives a better description 
of the high energy fusion data for $^{16}$O+$^{208}$Pb \cite{opb}.  
However, it was necessary to supplement the nuclear interaction 
with a weak, short-ranged imaginary potential.
It is interesting that the same conclusions apply to 
to the high-energy fusion of $^{16}$O+$^{16}$O.

\begin{figure}
\includegraphics[width = 11cm]{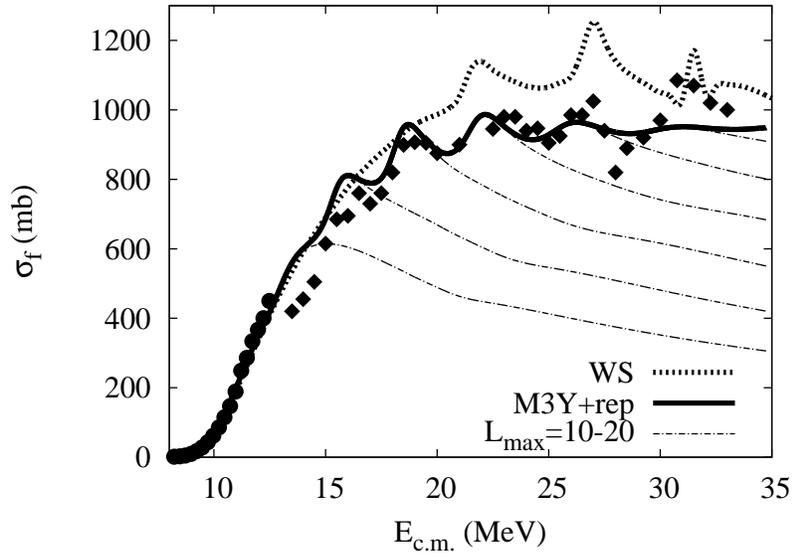}
\caption{\label{lfus} The measured fusion cross sections for 
$^{16}$O+$^{16}$O (solid circles \cite{Thomas}, diamonds: \cite{Tserruya})
and coupled-channels calculations that are based on the
Woods-Saxon (top dashed) and M3Y+repulsion potential 
(with $a_{\rm rep}=0.41$, solid curve).
The thin dashed curves show the dependence on 
the maximum angular momentum $L_{max}$.
All calculations include a short-range imaginary potential.}
\end{figure}

The results of coupled-channels calculations that are based
on the M3Y+repulsion potential and a short-ranged imaginary potential 
that acts near the minimum of the potential pocket are compared 
in Fig. \ref{lfus} to the data of Tserruya et al. \cite{Tserruya}. 
The data exhibit an oscillatory or step-wise increasing behavior 
which was also seen in the earlier data by Kolata et al. \cite{Kolata}
This behavior is qualitatively reproduced by the calculation
(solid curve).
The step-wise increase in the calculated cross section is
correlated with overcoming the potential barriers for $L$ = 12, 
14, 16, and 18. This can be seen by comparing to the thin dashed
curves which show the cross sections one obtains by imposing
different values of the maximum angular momentum $L_{max}$ in 
the calculations. Note that only even values of $L$ are considered 
for a symmetric system.

The coupled-channels calculations that are based on the deep 
Woods-Saxon potential are shown by the upper thick dashed curve 
in Fig. \ref{lfus}. It is seen that the data are suppressed 
compared to this calculation and that the step-wise behavior 
sets in at a higher energy and a higher angular momentum 
($L$=16 to be precise, compared to $L$=12 in the solid curve.)
Thus the high energy $^{16}$O+$^{16}$O fusion data show a 
clear preference for the shallow potential produced by the
M3Y+repulsion interaction.

The analysis of the elastic scattering data for $^{16}$O+$^{16}$O 
by Gobbi et al. \cite{gobbi} also revealed the need for a shallow 
potential.  The potential they obtained is illustrated by the upper 
thick dashed curve in Fig. \ref{pot}. 
The minimum of the pocket is in this case at 0.78 MeV, which is 
slightly  deeper that the 2.4 MeV pocket produced by the 
M3Y+repulsion potential (with $a_{\rm rep}$ = 0.41 fm.)
Thus it appears that both the elastic scattering data and the
high energy fusion data prefer a shallow pocket in the entrance
channel. 

It should be mentioned that the Gobbi potential does not provide 
a good description of the low-energy fusion data by Thomas et al.
\cite{Thomas}, although it has a shallow pocket. 
The reason is that the Coulomb barrier is not as thick as the one 
produced by the M3Y+repulsion potential (see Fig. \ref{pot}.)
As a consequence, the fusion data are hindered compared
to calculations that are based on the Gobbi potential.  

\section{Conclusions}

It has been shown that the calculated fusion cross sections
for $^{16}$O+$^{16}$O are sensitive to couplings to the $2^+$
and $3^-$ excited states of $^{16}$O even at low energies, 
where the excitation channels are closed.
Unfortunately, the overall quality of the fit to the fusion data
by Thomas et al. \cite{Thomas} 
is not very sensitive to the ion-ion potential at short distances
between the reacting nuclei. It is only at the very lowest energies 
that there is a preference for a shallow potential in the 
entrance channel.

The potential that gives the best fit to the fusion data by 
Thomas et al. \cite{Thomas} is 
the M3Y double-folding potential which has been corrected 
for the effect of the nuclear incompressibility.
This M3Y potential is calculated with a density
that is consistent with the measured charge radius of $^{16}$O,
and it produces a very realistic height of the Coulomb barrier.
The repulsive interaction that simulates the effect of the nuclear
incompressibility is calculated 
with parameters (the nuclear incompressibility and a diffuseness
parameter) that are similar to those that have been used previously
to reproduce the low-energy fusion data for medium-heavy systems.

The fusion cross sections obtained in coupled-channels calculations 
are in fairly good agreement with measurements at high energies when 
the calculations are based on the shallow M3Y+repulsion potential.  
In particular, the oscillatory or step-wise increasing behavior is 
reproduced very well, whereas the calculations that are based on the 
deeper Woods-Saxon do not reproduce the data.
The evidence for a shallow entrance channel potential is corroborated 
by the empirical optical potential for the elastic scattering obtained 
by Gobbi et al. \cite{gobbi}.

The $S$ factor obtained in the coupled-channels calculations that
give the best fit to the data by Thomas et al. has a maximum at a 
center-of-mass energy near 4 MeV.
The value of the $S$ factor at this energy is close to the value that 
was predicted more than 20 years ago in an adiabatic TDHF calculation. 
However, if the potential is adjusted to improve the fit only to the 
low-energy data, 
one obtains an even stronger hindrance of fusion at lower energies. 
This is in qualitative agreement with the empirical extrapolation 
proposed recently by Jiang et al.  \cite{cljastro}.
To confirm the hindrance experimentally one would have to measure
the fusion cross section down to an energy of 5-6 MeV.

{\bf Acknowledgments} 
The author is grateful to C. L. Jiang for many discussions. 
This work was supported by the U.S. Department of Energy, 
Office of Nuclear Physics, under Contract No. DE-AC02-06CH11357.

\begin{table}
\caption{Properties of the $2^+$ and $3^-$ states in $^{16}$O.
The B-values and Coulomb coupling strengths are from \cite{ENSDF}.
The nuclear couplings are from $\alpha$ scattering \cite{alfa}.}
\begin{ruledtabular}
\begin{tabular} {|c|c|c|c|c|c|c|}
Nucleus &
$\lambda^\pi$ &  E$_x$ (MeV) & B(E$\lambda$) (W.u.) & 
\ $\beta_\lambda^C$ & 
$(\frac{\beta R}{\sqrt{4\pi}})_C$ (fm) & 
$(\frac{\beta R}{\sqrt{4\pi}})_N$ (fm) \\
\colrule
$^{16}$O & $2^+$      & 6.92  &  3.1(1) & 0.35 & 0.30 & 0.27 \\
         & $3^-$      & 6.13  & 13.5(7) & 0.72 & 0.61 & 0.40 \\
\end{tabular}
\end{ruledtabular}
\end{table}

\end{document}